\documentclass[twocolumn,showpacs,preprintnumbers,amsmath,amssymb]{revtex4}
\usepackage{graphicx}
\usepackage{dcolumn}
\usepackage{color}
\usepackage{bm}
\begin{document}
\title {Charged oscillator in a heat bath in the presence of a magnetic field \& third law of thermodynamics.}
\author{ Malay Bandyopadhyay}
\affiliation{Department of Theoretical Physics, Tata Institute of Fundamental Research, Homi Bhabha Road, Colaba, Mumbai-400005, India.}
\vskip-2.8cm
\date{\today}
\vskip-0.9cm
\begin{abstract}
The quantum thermodynamic behaviour of a charged oscillator in the presence of a magnetic field and coupled to a heat bath through different coupling schemes is obtained analytically. It is shown that finite dissipation substitutes the zero-coupling result of exponential decay of entropy by a power law behaviour at low temperature. For the coordinate-coordinates coupling scheme the low temperature explicit results for the case of Ohmic, exponentially correlated and more generalized heat bath models are derived. In all the above mentioned cases free energy and entropy vanish linearly with temperature ($T$) as $T\rightarrow 0$ in conformity with Nernst's theorem. It is seen that coordinate (velocity)-velocities (coordinates) coupling is much more beneficial than the coordinate-coordinates coupling to ensure third law of thermodynamics. The case of radiation heat bath shows $T^3$ decay behaviour for entropy as $T\rightarrow 0$. It is observed that at low temperature free energy and entropy decay faster for the velocity-velocities scheme than  any other coupling schemes. This implies velocity-velocities coupling scheme is the most advantageous coupling scheme in restoring the third law of thermodynamics. It is shown that the low temperature thermodynamic functions are independent of magnetic field for all the above mentioned cases except the without dissipation case.  
\end{abstract}
\maketitle
{\section {Introduction}}
The third law of thermodynamics is an axiom of nature regarding entropy and the impossibility of reaching absolute zero of temperature. The third law was developed by Walther Nernst and is thus sometimes referred to as Nernst's theorem or Nernst's postulate \cite{a,b}. The third law of thermodynamics states that the entropy of a system at zero temperature is a well defined constant. This is because a system at zero temperature exists in its ground state, so that its entropy is determined only by the degeneracy of the ground state. The constant value of the entropy at absolute zero temperature is given by $ S = k_b \ln g $ (g=degeneracy). In the thermodynamic limit ($N\rightarrow\infty$) the typical value of entropy $S_0=\frac{S(T=0)}{N}$ goes to zero as long as the degeneracy does not grow with $N$ faster than exponentially \cite{c}. Thus it is impossible by any procedure, no matter however idealised, to reduce any system to the absolute zero of temperature in a finite number of operations.\\
A detail discussion on the history of the third law and the controversies can be found in the books by Dugdale \cite{d} and by Wilks \cite{e}. According to Max Planck the entropy per particle approaches a constant value $S_0$ at absolute zero temperature and it can generally be set equal to zero \cite{f}. Third law implies that thermodynamical quantities such as free energy, entropy, specific heat, isobar thermal coefficients etc. approach zero as $T\rightarrow 0$.\\
In this paper I investigate low temperature thermodynamic properties for open quantum system which is coupled to a heat bath by different coupling schemes. The recent development in the subject of quantum thermodynamics \cite{g,h,i} and widespread interest on the low temperature physics of small quantum systems has raised up the question : Does the third law of thermodynamics hold at quantum regime? How quantum dissipation can play an important role in thermodynamic theory? Recently P. Hanggi and G. L. Ingold have shown that finite dissipation actually helps to ensure the third law of thermodynamics \cite{j}. Further investigations has been made by W. C. Yang and B. J. Dong on the influence of various coupling forms \cite{k}. In this work I investigate the effect of different coupling schemes and different non-Markovian heat bath on the low temperature thermodynamical functions of a chrged dissipative oscillator in the presence of an external magnetic field.\\
With this preceding background I organize the rest of the paper as follows. In the next section I introduce the model system and different coupling schemes. In Sec. III, I discuss coordinate-coordinates coupling scheme. In this connection the without dissipation case is analyzed. In addition analytical explicit results of low temperature thermodynamical quantities for Ohmic, exponentially correlated, and arbitrary heat bath are derived. Coordinate (velocity)- velocities (coordinates) scheme is examined in details in Sec. IV. In this connection the radiation heat bath case is examined in details. Section V is for velocity-velocities coupling scheme. Finally I conclude this paper in Sec. VI.\\
{\section{Model System}}
The starting point of this section is the generalized Caldeira-Legget system-plus-reservoir Hamiltonian model for a charged particle $`e'$ in a magnetic field $\vec{B}$ in the operator form \cite{l,m} :
\begin{eqnarray}
\hat{H}&=&\frac{\Big(\hat{p}-e\vec{A}/c\Big)^2}{2m}+\frac{1}{2}m\omega_0^2\hat{r}^2 \nonumber \\
&&+\sum_{j=1}^N\Big\lbrack\frac{1}{2m_j}\Big(\hat{p}_j^2+m_j^2\omega_j^2\hat{q}_j^2\Big)+g(\hat{r},\hat{p},\hat{q}_j,\hat{p}_j)\Big\rbrack,
\end{eqnarray}
where $\lbrace\hat{r},\hat{p}\rbrace$ and $\lbrace\hat{q}_j,\hat{p}_j\rbrace$ are the sets of co-ordinate and momentum operators of system and bath oscillators. They follow the following commutation relations
\begin{eqnarray}
\lbrack \hat{r}_{\alpha},\hat{p}_{\beta}\rbrack = i\hbar\delta_{\alpha\beta},
\lbrack\hat{q}_{i\alpha},\hat{p}_{j\beta}\rbrack=i\hbar\delta_{ij}\delta_{\alpha\beta},
\end{eqnarray}
where $\alpha$, $\beta  $ denote components of the above mentioned operators along x, y directions. Equation (1) includes four types of bilinear couplings between the system and the environmental degrees of freedom. For the usual coordinate-coordinates coupling \cite{l}
\begin{equation}
g=-c_j\hat{r}\hat{q}_j+\frac{c_j^2\hat{r}^2}{2m_j\omega_j^2},
\end{equation}
for the system coordinate and environmental velocities coupling \cite{m}
\begin{equation}
g=-d_{1,j}\frac{\hat{r}\hat{p}_j}{m_j}+\frac{d_{1,j}^2\hat{r}^2}{2m_j},
\end{equation}
or for the system velocity and environmental coordinates coupling \cite{n}
\begin{equation}
g=-d_{2,j}\frac{\hat{p}\hat{q}_j}{m}+\frac{d_{2,j}^2\hat{q}_j^2}{2m},
\end{equation} 
and finally for system velocity and environmental velocities coupling \cite{o}
\begin{equation}
g=-e_{j}\frac{\hat{p}\hat{p}_j}{mm_j}+\frac{e_{j}^2\hat{p}_j^2}{2mm_j^2}.
\end{equation}
The additional terms appearing in the coupling are in order to compensate coupling induced potential and mass renormalization.\\
Eliminating the bath degrees of freedom by the Heisenberg equations of motion one can obtain the generalized quantum Langevin equation (GQLE) \cite{p,q}
\begin{eqnarray}
m\ddot{\hat{r}}+\int_{-\infty}^tdt^{\prime}\gamma_{\mu}(t-t^{\prime})\dot{\hat{r}}(t^{\prime})-\frac{e}{c}\dot{\hat{r}}\times\vec{B}+\vec{\nabla}V(\hat{r})=\hat{\theta}(t),
\end{eqnarray}
where dot denotes differentiation with respect to time t and $\mu=1,2,3,4$ is the variable for four different coupling schemes. The effect of the magnetic field is solely represented by the quantum version of the Lorentz force (third term in Eq. (7)). The memory kernel $\gamma_{\mu}(t)$ and the operator valued random force $\hat{\theta}$(t) are unaffected by the magnetic field. In this work I consider the confining potential to be harmonic for which an exact analysis is possible. It  is now possible to represent nonequal time anticommutator and commutator of $\hat{\theta}(t)$ as follows:
\begin{eqnarray}
\langle\lbrace\theta_{\alpha}(t),\theta_{\beta}(t^{\prime})\rbrace\rangle&=&\delta_{\alpha\beta}\frac{\beta\hbar}{\pi}\int_0^{\infty}d\omega J_{\mu}(\omega)\coth\Big(\frac{\beta\hbar\omega}{2}\Big)\nonumber \\ 
&&\times \cos\lbrack\omega(t-t^{\prime})\rbrack,\\
\hskip-0.5cm
\langle\lbrack\theta_{\alpha}(t),\theta_{\beta}(t^{\prime})\rbrack\rangle&=&\delta_{\alpha\beta}\frac{\beta\hbar}{\pi}\int_0^{\infty}d\omega J_{\mu}(\omega)\sin\omega(t-t^{\prime}),
\end{eqnarray}
where $\beta=\frac{1}{k_BT}$ is the inverse temperature.The memory kernel is given by
\begin{equation}
\gamma_{\mu}(t)=\frac{2}{m\pi}\int_0^{\infty}d\omega \frac{J_{\mu}(\omega)}{\omega}\cos(\omega t).
\end{equation} 
In equations (8)-(10) $J_{\mu}(\omega)$ denotes the spectral density function of the heat bath oscillators for different types of coupling schemes and is given as follows :
\begin{eqnarray}
J_1(\omega)=J_{c-c}(\omega)=\pi\sum_{j=1}^N\frac{c_j^2}{2m_j\omega_j}\delta(\omega-\omega_j),\\
J_2(\omega)=J_{c-v}(\omega)=\pi\sum_{j=1}^N\frac{d_{1,j}^2}{2m_j}\omega_j\delta(\omega-\omega_j),\\
J_3(\omega)=J_{v-c}(\omega)=\pi\sum_{j=1}^N\frac{d_{2,j}^2}{2m_j}\omega_j\delta(\omega-\omega_j),\\
J_4(\omega)=J_{v-c}(\omega)=\pi\sum_{j=1}^N\frac{e_j^2}{2m_j}\omega_j^3\delta(\omega-\omega_j).
\end{eqnarray}
I am interested in investigating low temperature thermodynamic behaviour of the model system described above. One can easily determine the free energy for this model system by using the remarkable formula \cite{r,s}
\begin{eqnarray}
F(T,B)=\frac{1}{\pi}\int_0^{\infty}d\omega f(\omega,T)\Im\Big\lbrack\frac{d}{d\omega}\ln\Big(\det\alpha(\omega+i0^+)\Big)\Big\rbrack,
\end{eqnarray}
where $f(\omega,T)$ is the free energy of a single oscillator of frequency $\omega$ and is given by 
\begin{equation}
f(\omega,T)=k_BT\log\Big\lbrack 1 - \exp(-\frac{\hbar\omega}{k_BT})\Big\rbrack,
\end{equation}
where I have ignored the zero-point contribution as I am interested in finite but low teperature effect. Here $\alpha(\omega)$ denotes the generalized susceptibility of the model system. Now I can rewrite Eq. (15) as follows \cite{r,s}:
\begin{equation}
F(T,B)=F(T,0)+\Delta F(T,B),
\end{equation}
where 
\begin{equation}
F(T,0)=\frac{3}{\pi}\int_0^{\infty}d\omega f(\omega, T)\Im\Big\lbrack\frac{d}{d\omega}\ln\alpha^{(0)}(\omega)\Big\rbrack
\end{equation}
is the free energy of the oscillator in the absence of the magnetic field and the correction due to the magnetic field is given by
\begin{equation}
\Delta F(T,B)=-\frac{1}{\pi}\int_0^{\infty}d\omega f(\omega, T)\Im\Big\lbrace\frac{d}{d\omega}\ln\Big\lbrack 1-\Big(\frac{eB\omega\alpha^{(0)}}{c}\Big)^2\Big\rbrack\Big\rbrace,
\end{equation}
where $\alpha^{(0)}(\omega)$ is the scalar susceptibility in the absence of a magnetic field. The function $f(\omega,T)$ vanishes exponentially for $\omega>>\frac{k_BT}{\hbar}$ and hence all the integrand in Eq. (18) and in Eq. (19) are confined to low frequencies. Thus one can easily obtain an explicit results by expanding the factor multiplying $f(\omega, T)$ in powers of $\omega$.\\
The scalar susceptibility for a harmonic oscillator in the absence of a magnetic field is given by \cite{r,s}
\begin{equation}
\alpha^{(0)}(\omega)=\frac{1}{m(\omega_0^2-\omega^2)-i\omega\tilde{\gamma}(\omega)},
\end{equation}
where
\begin{equation}
\tilde{\gamma}(\omega)=\int_0^tdt^{\prime}\gamma(t^{\prime})e^{i\omega t^{\prime}}.
\end{equation}
Now I can calculate $I_1=\Im \Big\lbrack\frac{d}{d\omega}\ln\alpha^{(0)}(\omega)\Big\rbrack$ and $I_2=\Im\Big\lbrack\frac{d}{d\omega}\ln\lbrack 1-\Big(\frac{eB\omega}{c}\alpha^{0}(\omega)\Big)^2\rbrack\Big\rbrack$. The expressions are given as follows
\begin{equation}
I_1=\frac{m\tilde{\gamma}(\omega)(\omega_0^2+\omega^2)+m\omega\tilde{\gamma}^{\prime}(\omega)(\omega_0^2-\omega^2)}{\lbrack m^2(\omega_0^2-\omega^2)^2+\omega^2\tilde{\gamma}^2(\omega)\rbrack},
\end{equation}
and 
\begin{eqnarray}
\hskip-0.8cm
&&I_2=2\frac{m\tilde{\gamma}(\omega)(\omega_0^2+\omega^2)+m\omega\tilde{\gamma}^{\prime}(\omega)(\omega_0^2-\omega^2)}{\lbrack m^2(\omega_0^2-\omega^2)^2+\omega^2\tilde{\gamma}^2(\omega)\rbrack} \nonumber \\
\hskip-0.8cm
&&-\frac{m\tilde{\gamma}(\omega)(\omega_0^2+\omega^2)+m\omega\tilde{\gamma}^{\prime}(\omega)(\omega_0^2-\omega^2+\omega\omega_c)}{\lbrack m^2(\omega_0^2-\omega^2+\omega\omega_c)^2+\omega^2\tilde{\gamma}^2(\omega)\rbrack} \nonumber \\
\hskip-0.8cm
&&-\frac{m\tilde{\gamma}(\omega)(\omega_0^2+\omega^2)+m\omega\tilde{\gamma}^{\prime}(\omega)(\omega_0^2-\omega^2-\omega\omega_c)}{\lbrack m^2(\omega_0^2-\omega^2-\omega\omega_c)^2+\omega^2\tilde{\gamma}^2(\omega)\rbrack},
\end{eqnarray}
where $\omega_c=\frac{eB}{mc}$ is the cyclotron frequency. I have now all the essential ingredients to calculate thermodynamic functions. My main task is to find free energy $F$ using Eqs. (17), (18) and (19) at low temperature. Then one can easily derive other thermodynamic functions at low temperature. Entropy is defined as 
\begin{equation}
S=-\frac{\partial F}{\partial T}.
\end{equation}
 In the next section I discuss usual coordinate-coordinates coupling scheme. In this connection I analyze without dissipation case. In addition I examine the decay behaviour of entropy with temperature for Ohmic, exponentially correlated, and arbitrary heat bath models.\\
\section{Coordinate-coordinates coupling scheme}
First, let me consider the usual case of the system's coordinate coupled to the coordinates of the heat bath. This kind of coupling can be realized experimentally in the case of a RLC circuit driven by a Gaussian white noise.\\
\subsection{Without Dissipation}
The limit of without dissipation can easily be obtained by taking $\tilde{\gamma}(\omega)=0$. Thus 
\begin{equation}
\alpha^{(0)}(\omega)=-\frac{1}{m(\omega^2-\omega_0^2)},
\end{equation}
and
\begin{equation}
\Big\lbrack 1-\Big(\frac{eB\omega}{c}\Big)^2\lbrack\alpha^{(0)}(\omega)\rbrack^2\Big\rbrack=\frac{\Big\lbrack (\omega^2-\omega_0^2)^2-(\omega\omega_c)^2\Big\rbrack}{(\omega^2-\omega_0^2)^2},
\end{equation}
where $\omega_c=\frac{eB}{mc}$ is the cyclotron frequency. For this case 
\begin{equation}
I_1=\Im\Big\lbrace\frac{d}{d\omega}\ln\alpha^{(0)}(\omega)\Big\rbrace =\pi\Big\lbrack\delta(\omega-\omega_0)+\delta(\omega+\omega_0)\Big\rbrack,
\end{equation}
where I have used the identity
\begin{equation}
\frac{1}{\omega-\omega_j+i0^+}=P\Big\lbrack\frac{1}{\omega-\omega_j}\Big\rbrack-i\pi\delta(\omega-\omega_j).
\end{equation}
Thus
\begin{equation}
F(T,0)=3f(\omega_0,T)
\end{equation} 
Similarly one can show that
\begin{equation}
\Delta F(T,B)=f(\omega_1,T)+f(\omega_2,T)-2f(\omega_0,T),
\end{equation}
where $\omega_{1,2}=\pm\frac{\omega_c}{2}+\lbrack \omega_0^2+(\frac{\omega_c}{2})^2\rbrack^{\frac{1}{2}}$. Hence
\begin{equation}
F(T,B)=f(\omega_0,T)+f(\omega_1,T)+f(\omega_2,T).
\end{equation}
At low temperature free energy becomes
\begin{equation}
F(T,B)=-k_BT\Big(e^{-\frac{\hbar\omega_0}{k_BT}}+e^{-\frac{\hbar\omega_1}{k_BT}}+e^{-\frac{\hbar\omega_2}{k_BT}}\Big).
\end{equation}
Thus it can be concluded that S(T) vanishes exponentially when $T\rightarrow 0$ for my model system without the dissipative heat bath.\\
\subsection{Ohmic heat bath}
For pure Ohmic heat bath one can take $\tilde{\gamma}(\omega)=m\gamma$, where $\gamma$ is friction constant. Thus the response function in the absence of magnetic field becomes
\begin{equation}
\alpha^{(0)}(\omega)=\Big\lbrack m(\omega_0^2-\omega^2)-im\omega\gamma\Big\rbrack^{-1}.
\end{equation} 
Thus
\begin{eqnarray*}
I_1&=&\frac{\gamma(\omega^2+\omega_0^2)}{(\omega^2-\omega_0^2)^2+\gamma^2\omega^2}\\
&&\stackrel{\omega\rightarrow 0}{\simeq}\frac{\gamma}{\omega_0^2}.
\end{eqnarray*}
Similarly
\begin{eqnarray*}
\hskip-1.5cm
&&-I_2=\frac{\gamma(\omega^2+\omega_0^2)}{(\omega^2-\omega_0^2+\omega\omega_c)^2+\gamma^2\omega^2}  \\
&&+ \frac{\gamma(\omega^2+\omega_0^2)}{(\omega^2-\omega_0^2-\omega\omega_c)^2+\gamma^2\omega^2}  \\
&&-2\frac{\gamma(\omega^2+\omega_0^2)}{(\omega^2-\omega_0^2)^2+\gamma^2\omega^2} \\
&&\stackrel{\omega\rightarrow 0}{\simeq}\frac{\gamma}{\omega_0^2}+\frac{\gamma}{\omega_0^2}-2\frac{\gamma}{\omega_0^2} = 0.
\end{eqnarray*}
Hence free energy of the system at low temperature can be written as
\begin{equation}
F(T)=\frac{3k_BT\gamma}{\pi\omega_0^2}\int_0^{\infty}d\omega\ln\Big\lbrack 1-\exp\big(-\frac{\hbar\omega}{k_BT}\Big)\Big\rbrack.
\end{equation}
The following integral is relevent for my calculation throughout this paper: 
\begin{equation}
\int_0^{\infty}dy y^{\nu}\log(1-e^{-y})=-\Gamma(\nu+1)\zeta(\nu+2),
\end{equation}
where $\Gamma$ is gamma function and $\zeta$ is Riemann's zeta function.
Using this integral one can show
\begin{equation}
F(T)=-\frac{\pi}{2}\hbar\gamma\Big(\frac{k_BT}{\hbar\omega_0}\Big)^2.
\end{equation}
Hence entropy is given by
\begin{equation}
S(T)=\pi\hbar\gamma\frac{k_B^2T}{(\hbar\omega_0)^2}.
\end{equation}
As $T\rightarrow 0$, $S(T)$ vanishes linearly with $T$ which perfectly matches with third law of thermodynamics. It shows the usual Ohmic friction behaviour of linear decay.\\
Now if I consider that the environmental oscillators have a power spectrum with a narrow Lorentzian peak centred at a finite frequency not at zero. Thus the Fourier transform of the memory function is
\begin{equation}
\tilde{\gamma}(\omega)=\frac{m\gamma\Omega^4}{\Gamma^2\omega^2+(\Omega^2-\omega^2)^2},
\end{equation}
where $\gamma$ denotes the Markovian friction strength of the system, $\Gamma$ and $\Omega$ are the damping and frequency parameters of the harmonic noise \cite{t}.
\begin{eqnarray*}
&&I_1=\frac{m^2\gamma\Omega^4DA-m^2\gamma\Omega^4\omega A^{\prime}C}{m^2C^2A^2+4m^2\gamma^2\Omega^8\omega^2}\\
&&\stackrel{\omega\rightarrow 0}{\simeq}\frac{\gamma}{\omega_0^2}
\end{eqnarray*}
where $A=\lbrack\Gamma^2\omega^2+(\Omega^2-\omega^2)^2\Big\rbrack$, $A^{\prime}=\frac{dA}{d\omega}$, $C=(\omega_0^2-\omega^2)$ and $D=(\omega^2+\omega_0^2)$. Now 
\begin{eqnarray*}
-I_2&=&\frac{m^2\gamma\Omega^4DA-m^2\gamma\Omega^4\omega A^{\prime}C_1}{m^2C_1^2A^2+4m^2\gamma^2\Omega^8\omega^2}\\
&&+\frac{m^2\gamma\Omega^4DA-m^2\gamma\Omega^4\omega A^{\prime}C_2}{m^2C_2^2A^2+4m^2\gamma^2\Omega^8\omega^2}\\
&&-2\frac{m^2\gamma\Omega^4DA-m^2\gamma\Omega^4\omega A^{\prime}C}{m^2C^2A^2+4m^2\gamma^2\Omega^8\omega^2}\\
&&\stackrel{\omega\rightarrow 0}{\simeq}\frac{\gamma}{\omega_0^2}+\frac{\gamma}{\omega_0^2}-2\frac{\gamma}{\omega_0^2}= 0,
\end{eqnarray*}
where $C_1=(\omega^2-\omega_0^2+\omega\omega_c)$ and $C_2=(\omega^2-\omega_0^2-\omega\omega_c)$. Thus free energy of my model system for the heat bath with Lorentzian peak power spectrum is given by
\begin{equation}
F(T)= -\frac{\pi}{2}\hbar\gamma\Big(\frac{k_BT}{\hbar\omega_0}\Big)^2.
\end{equation}
Hence the decay behaviour of entropy is again the same as of Eq. (37).
\subsection{Exponentially correlated heat bath}
In this subsection I consider exponentially correlated heat bath whose memory friction is given by
\begin{equation}
\gamma(t)=\frac{m\gamma}{\tau_c}e^{-\frac{|t|}{\tau_c}}.
\end{equation}
The Fourier transform of memory friction gives us
\begin{equation}
\tilde{\gamma}(\omega)=\frac{m\gamma}{1+\omega^2\tau_c^2}.
\end{equation}
Now the required expressions for $I_1$ and $I_2$ are as follows
\begin{eqnarray*}
&&I_1=\frac{m^2\gamma(1+\omega^2\tau_c^2)D-2m^2\omega^2\gamma\tau_c^2C}{m^2C^2(1+\omega^2\tau_c^2)^2+m^2\gamma^2\omega^2}\\
&&\stackrel{\omega\rightarrow 0}{\simeq}\frac{\gamma}{\omega_0^2}.
\end{eqnarray*}
Similarly
\begin{eqnarray*}
\hskip-1.5cm
&&-I_2=\frac{m^2\gamma(1+\omega^2\tau_c^2)D-2m^2\omega^2\gamma\tau_c^2C_1}{m^2C_1^2(1+\omega^2\tau_c^2)^2+m^2\gamma^2\omega^2}\\
&&+\frac{m^2\gamma(1+\omega^2\tau_c^2)D-2m^2\omega^2\gamma\tau_c^2C_2}{m^2C_2^2(1+\omega^2\tau_c^2)^2+m^2\gamma^2\omega^2}\\
&&-2\frac{m^2\gamma(1+\omega^2\tau_c^2)D-2m^2\omega^2\gamma\tau_c^2C}{m^2C^2(1+\omega^2\tau_c^2)^2+m^2\gamma^2\omega^2}\\
&&\stackrel{\omega\rightarrow 0}{\simeq}\frac{\gamma}{\omega_0^2}+\frac{\gamma}{\omega_0^2}-2\frac{\gamma}{\omega_0^2} = 0.
\end{eqnarray*}
Thus free energy of this model system with exponentially correlated heat bath is given by
\begin{equation}
F(T)= -\frac{\pi}{2}\hbar\gamma\Big(\frac{k_BT}{\hbar\omega_0}\Big)^2.
\end{equation}
Hence the decay behaviour of entropy with temperature is again linear which is same as that of Ohmic friction results.
\subsection{Arbitrary heat bath}
The heat bath is characterized by the memory friction function $\tilde{\gamma}(z)$. According to Ford {\it et al} \cite{i} it should be positive and must be analytic in the upper half plane and must satisfy the reality condition
\begin{equation}
\tilde{\gamma}(-\omega+i0^+)=\tilde{\gamma}(\omega+i0^+).
\end{equation}
Now according to Ford {\it et al} \cite{i} this must be in the neighbourhood of origin as follows :
\begin{equation}
\tilde{\gamma}(\omega)\simeq m b^{1-\nu}(-i\omega)^{\nu},
\end{equation}
where $-1<\nu<1$, and $b$ is a positive constant with dimensions of frequency. Thus the scalar susceptibility of the model system in the absence of the magnetic field is given by
\begin{equation}
\alpha^{(0)}(\omega)=\frac{1}{m(\omega_0^2-\omega^2)+mb^{1-\nu}(-i\omega)^{1+\nu}}.
\end{equation}
Thus 
\begin{eqnarray*}
I_1&=&\frac{b^{1-\nu}\omega^{\nu}\cos(\frac{\nu\pi}{2})\Big\lbrack(1+\nu)C+2\omega^2\Big\rbrack}{\Big|C+b^{1-\nu}(-i\omega)^{1+\nu}\Big|^2}\\
&&\stackrel{\omega\rightarrow 0}{\simeq}(1+\nu)\cos(\frac{\nu\pi}{2})\frac{b^{1-\nu}}{\omega_0^2}\omega^{\nu},
\end{eqnarray*}
and 
\begin{eqnarray*}
&&-I_2=\frac{b^{1-\nu}\omega^{\nu}\cos(\frac{\nu\pi}{2})\Big\lbrack(1+\nu)C_1+2\omega^2\Big\rbrack}{\Big|C_1+b^{1-\nu}(-i\omega)^{1+\nu}\Big|^2}\\
&&+\frac{b^{1-\nu}\omega^{\nu}\cos(\frac{\nu\pi}{2})\Big\lbrack(1+\nu)C_2+2\omega^2\Big\rbrack}{\Big|C_2+b^{1-\nu}(-i\omega)^{1+\nu}\Big|^2}\\
&&-2\frac{b^{1-\nu}\omega^{\nu}\cos(\frac{\nu\pi}{2})\Big\lbrack(1+\nu)C+2\omega^2\Big\rbrack}{\Big|C+b^{1-\nu}(-i\omega)^{1+\nu}\Big|^2}\\
&&\stackrel{\omega\rightarrow 0}{\simeq}2(1+\nu)\cos(\frac{\nu\pi}{2})\frac{b^{1-\nu}}{\omega_0^2}\omega^{\nu}\\
&&-2(1+\nu)\cos(\frac{\nu\pi}{2})\frac{b^{1-\nu}}{\omega_0^2}\omega^{\nu}\\
&&=0.
\end{eqnarray*}
Hence free energy of the model system with such arbitrary heat bath is given by
\begin{equation}
F(T)=-3\Gamma(\nu+2)\zeta(\nu+2)\cos\Big(\frac{\nu\pi}{2}\Big)\frac{\hbar b^3}{\pi\omega_0^2}\Big(\frac{k_BT}{\hbar b}\Big)^{2+\nu}.
\end{equation}
Finally entropy of the system is 
\begin{equation}
S(T)=3\Gamma(\nu+3)\zeta(\nu+2)\cos\Big(\frac{\nu\pi}{2}\Big)\frac{k_Bb^2}{\pi\omega_0^2}\Big(\frac{k_BT}{\hbar b}\Big)^{1+\nu}.
\end{equation}
Since $(\nu+1)$ is always positive and hence $S(T)\rightarrow 0$ as $T\rightarrow 0$.
\section{Coordinate (Velocity) - Velocities (Coordinates) Coupling}
In this section I consider the typical case of the first kind where coordinate (velocity) of the system coupled to the velocities (coordinates) of the heat bath. The physical situations for such scheme can be found for a vortex transport in the presence of magnetic field \cite{u} and particle interacting via dipolar coupling \cite{v}. The friction memory function for this kind of coupling scheme is given by
\begin{equation}
\gamma(t)=m\gamma\Gamma\exp\Big(-\frac{\Gamma t}{2}\Big)\Big(\cos(\omega_1t)-\frac{\Gamma}{2\omega_1}\sin(\omega_1 t)\Big).
\end{equation}
where $\omega_1^2=\Omega^2-\frac{\Gamma^2}{4}$, $\gamma$ is the Markovian friction strength, $\Gamma$ and $\Omega$ are the damping and frequency parameters of the harmonic noise. The Fourier transform of the memory friction function is given by
\begin{equation}
\tilde{\gamma}(\omega)=\frac{2m\gamma\Gamma^2\omega^2}{\Gamma^2\omega^2+(\Omega^2-\omega^2)^2}.
\end{equation}
The expressions of $I_1$ and $I_2$ for this particular scheme are given by 
\begin{eqnarray*}
&&I_1=\frac{2m^2\gamma\Gamma^2\omega^2DA+4m^2\gamma\Gamma^2\omega^2AC-2m^2\gamma\Gamma^2\omega^3A^{\prime}C}{m^2(AC)^2+4m^2\gamma^2\Gamma^4\omega^6}\\
&&\stackrel{\omega\rightarrow 0}{\simeq}\frac{6\gamma\Gamma^2}{\Omega^4\omega_0^2},
\end{eqnarray*}
and
\begin{eqnarray*}
&&-I_2=\frac{2m^2\gamma\Gamma^2\omega^2DA+4m^2\gamma\Gamma^2\omega^2AC_1-2m^2\gamma\Gamma^2\omega^3A^{\prime}C_1}{m^2(AC_1)^2+4m^2\gamma^2\Gamma^4\omega^6}\\
&&+\frac{2m^2\gamma\Gamma^2\omega^2 DA+4m^2\gamma\Gamma^2\omega^2AC_2-2m^2\gamma\Gamma^2\omega^3A^{\prime}C_2}{m^2(AC_2)^2+4m^2\gamma^2\Gamma^4\omega^6}\\
&&-2\frac{2m^2\gamma\Gamma^2\omega^2DA+4m^2\gamma\Gamma^2\omega^2AC-2m^2\gamma\Gamma^2\omega^3A^{\prime}C}{m^2(AC)^2+4m^2\gamma^2\Gamma^4\omega^6}\\
&&\stackrel{\omega\rightarrow 0}{\simeq}\frac{6\gamma\Gamma^2}{\Omega^4\omega_0^2}+\frac{6\gamma\Gamma^2}{\Omega^4\omega_0^2}-12\frac{6\gamma\Gamma^2}{\Omega^4\omega_0^2}=0.
\end{eqnarray*}
Thus free energy of the system becomes
\begin{equation}
F(T)=-\frac{2\gamma\Gamma^2\pi^3}{5\Omega^4\omega_0^2}\hbar\omega_0^2\Big(\frac{k_BT}{\hbar\omega_0}\Big)^4.
\end{equation}
Entropy is given by
\begin{equation}
S(T)=\frac{8\gamma\Gamma^2\pi^3}{5\omega_0^2\Omega^4}k_B\omega_0^2\Big(\frac{k_BT}{\hbar\omega_0}\Big)^3.
\end{equation}
Thus the decay behaviour of the entropy is much faster than that of the coordinate-coordinates coupling. This is due to the much stronger dependence of the memory friction kernel on the frequency and hence it changes the thermodynamic behaviour of the system a lot at low temperature.\\
\subsection{Radiation Heat Bath}
In this case the Fourier transform of the associated memory friction function is given by \cite{w} 
\begin{equation}
\tilde{\gamma}(\omega)=\frac{2e^2\omega\Omega^{\prime 2}}{3c^3(\omega+i\Omega^{\prime})},
\end{equation}
where $e$ is the charge of the radiation field, $c$ is the velocity of light, $\Omega^{\prime}$ is the large cut-off frequency. Thus
\begin{eqnarray*}
I_1&=&\frac{3\omega_0^2\tau_e\omega^2+\tau_e^3\omega_0^2\omega^4-\tau_e\omega^4}{\lbrack (\omega_0-\omega^2)^2+\omega^2\omega_0^4\tau_e^2\rbrack(1+\omega^2\tau_e^2)}\\
&&\stackrel{\omega\rightarrow 0}{\simeq}\frac{3\tau_e}{\omega_0^2}\omega^2,
\end{eqnarray*}
where $\tau_e=\frac{2e^2}{3Mc^3}$ and $M=m+\frac{2e^2\Omega^{\prime}}{3c^3}$=renormalized mass. Similarly 
\begin{eqnarray*}
-I_2&=&\frac{3\omega_0^2\tau_e\omega^2+\tau_e^3\omega_0^2\omega^4-\tau_e\omega^4}{\lbrack (\omega_0-\omega^2+\omega\omega_c)^2+\omega^2\omega_0^4\tau_e^2\rbrack(1+\omega^2\tau_e^2)}\\
&&+\frac{3\omega_0^2\tau_e\omega^2+\tau_e^3\omega_0^2\omega^4-\tau_e\omega^4}{\lbrack (\omega_0-\omega^2-\omega\omega_c)^2+\omega^2\omega_0^4\tau_e^2\rbrack(1+\omega^2\tau_e^2)}\\
&&-2\frac{3\omega_0^2\tau_e\omega^2+\tau_e^3\omega_0^2\omega^4-\tau_e\omega^4}{\lbrack (\omega_0-\omega^2)^2+\omega^2\omega_0^4\tau_e^2\rbrack(1+\omega^2\tau_e^2)}\\
&&\stackrel{\omega\rightarrow 0}{\simeq}\frac{6\tau_e}{\omega_0^2}\omega^2-\frac{6\tau_e}{\omega_0^2}\omega^2=0.
\end{eqnarray*}
Thus the free energy for the model system with radiation heat bath is given by
\begin{equation}
F(T)=-\frac{\pi^3}{5}\hbar\omega_0^2\tau_e\Big(\frac{k_BT}{\hbar\omega_0}\Big)^4.
\end{equation}
The decay behaviour of the entropy is given by the following expression :
\begin{equation}
S(T)=\frac{4\pi^3}{5}k_B\omega_0\tau_e\Big(\frac{k_BT}{\hbar\omega_0}\Big)^3.
\end{equation}
Comparing equation (51) with equation (54) one can concludes that the decay behaviour of entropy with temperature for the radiation heat bath is same as that of the coordinate (velocity)- velocities (coordinates) coupling scheme. This decay behaviour is faster than the coordinate-coordinates coupling scheme decay behaviour of the entropy with temperature.\\
\section{Velocity-Velocities Coupling}
In this section I consider the second atypical case of dissipation where the system's velocity is coupled to the velocities of the heat bath. This velocity-velocities coupling scheme practically exists in electromagnetic problems such as superconducting quantum interference devices \cite{c,j} or in electromagnetic field \cite{w}. The spectrum of the friction memory function is completely different from that of coordinate-coordinates coupling and velocity (coordinate)-coordinates (velocities) scheme. The Fourier transform of the memory friction function is given by
\begin{equation}
\tilde{\gamma}(\omega)=\frac{2m\gamma\omega^4}{\Gamma^2\omega^2+(\Omega^2-\omega^2)^2}.
\end{equation} 
The expressions for $I_1$ and $I_2$ for this particular scheme are as follows
\begin{eqnarray*}
I_1&=&\frac{2m^2\gamma\omega^4DA+8m^2\gamma\omega^4AC-2m^2\gamma\omega^5A^{\prime}C}{m^2(CA)^2+4m^2\gamma^2\omega^{10}}\\
&&\stackrel{\omega\rightarrow 0}{\simeq}\frac{10\gamma\omega^4}{\omega_0^2\Omega^4}.
\end{eqnarray*}
and
\begin{eqnarray*}
-I_2&=&\frac{2m^2\gamma\omega^4DA+8m^2\gamma\omega^4AC_1-2m^2\gamma\omega^5A^{\prime}C_1}{m^2(C_1A)^2+4m^2\gamma^2\omega^{10}}\\
&&+\frac{2m^2\gamma\omega^4DA+8m^2\gamma\omega^4AC_1-2m^2\gamma\omega^5A^{\prime}C_1}{m^2(C_2A)^2+4m^2\gamma^2\omega^{10}}\\
&&-2\frac{2m^2\gamma\omega^4DA+8m^2\gamma\omega^4AC-2m^2\gamma\omega^5A^{\prime}C}{m^2C^2A^2+4m^2\gamma^2\omega^{10}}\\
&&\stackrel{\omega\rightarrow 0}{\simeq}\frac{10\gamma\omega^4}{\omega_0^2\Omega^4}+\frac{10\gamma\omega^4}{\omega_0^2\Omega^4}-2\frac{10\gamma\omega^4}{\omega_0^2\Omega^4}=0.
\end{eqnarray*}
Thus the free energy of the charged oscillator in the magnetic field for velocity-velocities coupling heat bath is given by
\begin{equation}
F(T)=-\frac{144\pi^5}{189}\Big(\frac{\omega_0}{\Omega}\Big)^4\hbar\gamma\Big(\frac{k_BT}{\hbar\omega_0}\Big)^6.
\end{equation}
The decay behaviour of entropy with temperature follows :
\begin{equation}
S(T)=\frac{864\pi^5}{189\Omega}\Big(\frac{\omega_0}{\Omega}\Big)^3k_B\gamma\Big(\frac{k_BT}{\hbar\omega_0}\Big)^5.
\end{equation} 
Again as $T\rightarrow0$, entropy $S(T)\rightarrow 0$ in conformity with Nernst's theorem. But the decaying behaviour of $S(T)$ is even faster than the coordinate(velocity)-velocities(coordinates) coupling scheme. So velocity-velocities coupling scheme is the most beneficial for the validification of third law of thermodynamics.
\section{Summary \& Conclusions}
In this work I analyze the low temperature thermodynamic behaviour of a charged harmonic oscillator in the presence of an external magnetic field and is coupled with a heat bath through different coupling schemes. The free energy of the charged quantum harmonic oscillator in an arbitrary heat bath is given by a ``remarkable" formula of Ford et al \cite{r,s} which involves only a single integral. One can exactly calculate this integral at low temperature limit and hence the low tempearature thermodynamic properties are derived. Mainly the decay behaviour of entropy with temperature is studied for the charged oscillator system. \\
The presence of finite quantum dissipation changes the well known Einstein like behaviour of exponential decay of entropy into a power law behaviour. The case of an usual Ohmic heat bath, the case of an Ohmic heat bath with Lorentzian peak power spectrum, exponentially correlated heat bath and more generalized heat bath with coordinate - coordinates coupling  are discussed in details. In all the cases entropy vanishes at zero temperature in conformity with third law of thermodynamics. Ohmic heat bath and exponentially correlated heat bath shows the linear decay of entropy with temperature at very low temperature which is in accordance with the usual case of Ohmic friction. The decay behaviour of the low temperature thermodynamical functions for anomalous coupling cases are examined in details. In the first atypical case i.e. coordinate (velocity)- velocities (coordinates) case the entropy decays much faster ($T^3$) than the usual coordinate-coordinates coupling form when temperature approaches zero. In the case of radiation heat bath it is shown explicitly that entropy follows $T^3$ variation at low temperature. In the second anomalous case i.e. the velocity-velocities coupling entropy decays faster $(T^5)$ than any other coupling schemes as temperature approches zero. It is the most benifitial coupling scheme in the sense of validification of third law. Also one can note that low temperature thermodynamic functions are independent of $B$ in all the coupling schemes.\\
The results obtained for the low-temperature behaviour of thermodynamic functions of a charged oscillator in the presence of the magnetic field are not only of theoretical interest but it can be found to be relevent for experiments in nanosciences where one wants to examine the validity of quantum thermodynamics of small systems which are coupled to heat bath \cite{x}.  
\section*{Acknowledgments}
MB acknowledges financial support from the Department of Atomic Energy (DAE), Government of India.\\

\end{document}